\documentclass[12pt]{iopart}
\usepackage{amsfonts}
\usepackage{graphicx}
\usepackage{amsmath}

\usepackage[colorlinks]{hyperref}
\hypersetup{
    linkcolor=black,       
    citecolor=black,       
    filecolor=black,       
    urlcolor=black         
}

\begin{document}

\title[The effect of plasma expansion on the dispersion properties of MHD waves]{The effect of plasma expansion on the dispersion properties of MHD waves}
\author{Sebastián Saldivia$^1$, Felipe A. Asenjo$^2$, Pablo S. Moya$^1$}
\address{$^1$ Departamento de Física, Facultad de Ciencias, Universidad de Chile, Las Palmeras
3425, Santiago, 7800003, Chile}
\address{$^2$ Facultad de Ingeniería y Ciencias, Universidad Adolfo Ibáñez, Santiago 7491169, Chile.}
\ead{sebastian.saldivia@ug.uchile.cl, felipe.asenjo@uai.cl, pablo.moya@uchile.cl}
\begin{abstract}
In this work, we employ the set of ideal expanding magnetohydrodynamic (MHD) equations within the Expanding Box Model (EBM) framework to theoretically characterize the effects of radial solar wind expansion on its characteristic linear MHD waves. Through the analytical derivation of dispersion relations by a first-order expansion of the MHD-EBM equations, we explore the changes in wave propagation across a range of heliocentric distances on the linear magnetohydrodynamic modes: the Alfvén mode and the fast and slow magnetosonic modes, as obtained from the ideal MHD-EBM equations. Our findings reveal a spatial dependence in the derived dispersion relations that aligns with both the literature and the traditional ideal MHD case in the non-expanding limit, thereby helping to bridge the gap between theory and observation in solar wind dynamics. We observe a general decrease in wave frequencies as the plasma expands farther from the Sun. This decrease is reflected in the dispersion relations through the radial decrease of both the Alfvén and sound speeds, which decrease proportionally to $1/R$ and $1/R^{\gamma - 1}$, respectively, where $\gamma$ is the plasma polytropic index. The fast magnetosonic mode frequency and phase speed are significantly affected by the polytropic index value. We consider three models for the polytropic index evolution in the expanding solar wind: a constant (quasi-adiabatic) case, a radially decreasing profile in the outer heliosphere, and a model incorporating thermodynamic heating effects. Notably, we find that in the case of a decreasing polytropic index, the fast magnetosonic mode experiences an acceleration in the distant heliosphere, highlighting the significant influence of expansion on solar wind dynamics.
\end{abstract}
\submitto{\PS}
\maketitle
\section{Introduction}

The solar wind can be described as a continuous flow of plasma that expands at supersonic speeds from the Sun to the outer edge of the heliosphere. As it propagates in interplanetary space, the solar wind transfers and converts energy across a great range of scales through various multi-scale mechanisms, where the interaction between propagating plasma waves and turbulent motions plays a prominent role~\cite{Bruno2013,Marino2023}. Several space missions over the past decades have provided a deep understanding of these processes and phenomena in the solar wind from 0.3 AU to the limits of the heliosphere\cite{Coleman1968,Marsch1982,Bruno2013}. Recently, measurements at inner heliocentric distances have been studied with the contemporary Parker Solar Probe and Solar Orbiter missions~\cite{Fox2015,Mller2020,Rivera2024,Brodiano2023}. Despite advances in modeling and observations, several fundamental problems within the solar wind remain unsolved. One such problem is the heating and acceleration of solar wind plasma as it expands through interplanetary space.

One of the most established models for incorporating solar wind expansion in its thermodynamic evolution is the double-adiabatic or Chew–Goldberger–Low (CGL) theory~\cite{CGL}. In this model, the resulting magnetohydrodynamic (MHD) equations predict the evolution of the proton temperature anisotropy as inversely proportional to the parallel beta of the plasma, thus predicting an adiabatic cooling of the solar wind. Nevertheless, in situ observations from 0.3 to 1 AU have shown that the solar wind does not cool adiabatically during its expansion. In fact, it evolves more slowly than predicted by the CGL theory~\cite{Marsch1982,Matteini2007,Hellinger2011}. Furthermore, observations reveal that this deviation is not uniform across the well-known solar wind structures, which distinguishes between fast and slow streams. Fast streams cool more slowly than slow currents, with a stronger contrast for protons than electrons~\cite{Hellinger2011,Shi2022,Dakeyo2022}. This suggests the presence of heating mechanisms acting on the solar wind after it leaves the solar corona, which are still poorly understood.

Several physical mechanisms have been proposed to provide the energy necessary to heat the plasma to the observed temperatures. These include the turbulent cascade~\cite{Matthaeus2011,Bruno2013} and interaction with ions from interstellar space~\cite{Fahr2002,Zank2018}. In this context, solar wind heating by electromagnetic waves has been considered as a suitable scenario. Recent models have shown that during solar wind generation, Alfvénic waves may provide the necessary energy for plasma heating in fast solar wind, highlighting the role of minor ions (see e.g.~\cite{Ofman2007,Ofman2011,Moya2014} and references therein). Throughout the solar wind expansion, magnetohydrodynamic waves, mainly Alfvén waves, which propagate along the solar wind as it expands, are thought to play an important role in the heating of the solar wind, with a larger energy contribution to the dynamics of fast solar wind~\cite{Belcher1971b,Rville2020}. These MHD waves can be characterized by solving the MHD equations for homogeneous, infinite-extent plasma with plane-wave solutions. The three permitted decoupled waves correspond to the Alfvén wave, the fast magnetosonic wave, and the slow magnetosonic wave. These waves have been extensively studied and have well-defined characteristics~\cite{Tu1995,Chelpanov2022review}. Alfvén waves are transverse and propagate along or oblique to magnetic field lines, while magnetosonic waves are longitudinal and also propagate along magnetic field lines.

The evolution of MHD waves propagating in the fast solar wind and their contribution to solar wind heating have been a relevant research topic in recent decades. The first theoretical attempt to model this phenomenon was by Parker~\cite{Parker1965}, who studied the radial propagation of low-frequency Alfvén and magnetosonic waves in a spherically symmetric background magnetic field. Subsequently, Hollweg~\cite{Hollweg1973a} investigated the propagation of Alfvén waves in a two-fluid model of the solar wind. Later, Barnes and Hollweg~\cite{Barnes1974} developed linear WKB solutions for large-amplitude Alfvén waves in the solar wind. Isenberga~\cite{Isenberg1984} employed this linear analysis to study the Alfvén wave propagation in a multi-ion plasma, while later Hollweg~\cite{Hollweg1990} studied the 'toroidal' Alfvén waves in the solar wind made in the past, improving over the last results. In the following decade, Vasquez and Hollweg~\cite{Vasquez2004} studied the non-linear case for outgoing and ingoing MHD waves. More recently, Rivera et al.~\cite{Rivera2024} studied the evolution of a solar wind stream in the inner heliosphere. They found that the kinetic and thermal energy gained by the stream coincides with the energy lost by the observed large-amplitude Alfvén waves, thus showing that these waves can drive the fast solar wind’s heating and acceleration. As we can see, understanding the dynamics and propagation of Alfvén waves in the solar wind has been a major focus of study over the decades. An important aspect is that these waves propagate and interact in the solar wind as it expands over heliocentric distances. MHD waves are believed to play a significant role in solar wind heating, along with several other mechanisms. Therefore, it is important to understand how solar wind expansion affects the propagation and dispersion of MHD waves to address the broader problem of solar wind heating.

In this context, Velli et al.~\cite{Velli1992} developed an innovative approach to explicitly study the effect of plasma expansion on the dynamics and heating of the solar wind, known as the Expanding Box Model (EBM). The EBM allows the study of the radial expansion of the plasma in a non-inertial reference frame that moves along with the plasma parcel at constant velocity. This model describes the spherical expansion of the plasma through an approximation to a Cartesian coordinate system. In this system, the plasma box maintains a constant volume by renormalizing the transverse coordinates to the solar wind's propagation direction. This renormalization of the coordinates means that the plasma expansion is no longer a spatial property of the system. Instead, it introduces time variations in the MHD equations, which are translated into non-inertial forces modifying the conservation of macroscopic plasma quantities~\cite{Grappin1993,Grappin1996}. This has various analytical advantages, as the MHD equations in this formalism show an explicit dependence on the expanding parameters. Keeping the volume of the plasma box constant has important implications when performing numerical simulations, as it limits possible memory problems when simulating an expanding plasma system. 

The EBM has motivated an extensive history of research into the role of expansion in plasma heating dynamics, with research in simulations and theory spanning a range of scales from the microscopic to the macroscopic ~\cite{Innocenti_2019,Innocenti_2020,Micera2020,Echeverria-Veas2024}. Liewer et al.~\cite{Liewer2001} studied the dynamics of Alfvén and ion-cyclotron waves in the expanding solar wind through hybrid simulations, while Ofman et al.~\cite{Ofman2011} and Moya et al.~\cite{Moya2012} performed hybrid simulations to study the kinetics of protons and heavy ions through the evolution of their temperature anisotropy. In more recent research, the model has been generalized by considering an accelerating co-moving frame, allowing for the solar wind regions at heliocentric distances close to the Sun~\cite{Tenerani2017}, and extended to a quasilinear model of the expanding solar wind~\cite{Seough2023}. The propagation of Alfvén waves in the expanding solar wind has been studied through the EBM formalism, both in theory and simulations. In particular, Nariyuki~\cite{Nariyuki2015} explored solutions of the derivative nonlinear Schrödinger equation for nonlinear Alfvén waves in an accelerating EBM, and more recently, Shi et al.~\cite{Shi2020} carried out simulations of Alfvénic fluctuations propagating along the solar wind with a fast-slow stream interaction.
    
As mentioned, there are several examples in the literature of using EBM to study solar wind heating mechanisms. However, no analytical effort has yet been made to characterize the dispersive properties of linear MHD waves propagating in the solar wind within the EBM framework. Even though the EBM has analytical advantages by providing a system of MHD equations in the presence of non-inertial forces with an explicit dependence on the expansion~\cite{Grappin1993,Nariyuki2015,Echeverria-Veas2024}, to date, no theoretical analysis of the properties of linear MHD waves in the EBM frame has been done. This work uses these equations to derive the dispersion relations of the three normal modes of ideal MHD within the MHD-EBM framework. This allows us to analyze the effects and time dependencies induced by plasma expansion on Alfvén and magnetosonic waves. Our study focuses on small-amplitude wave propagation in an ideal fluid, thus examining the linear MHD waves case. Thus, we must pay special attention to the right-hand side of the MHD-EBM equations, as they represent anisotropic energy losses due to the inertial forces driving the expansion in the co-moving frame. We expect to be able to quantify the radial decay of wave frequency during plasma expansion due to the incorporation of these terms into the MHD equations, as they will propagate during the analytical derivation of the dispersion relations. One advantage of studying linear MHD waves in this expanding framework is that they have been extensively studied in a non-expanding context, allowing us to compare our results with the well-known literature. Therefore, in the following sections, we will perform a linearization of the MHD-EBM equations, where we derive the expanding dispersion relations.

This paper will be organized as follows: in Section \ref{sec2}, we introduce the EBM formalism and the modifications that are introduced by the expansion in the ideal MHD system of equations (MHD-EBM). In Section \ref{sec3}, we develop the theoretical derivation for the wave equation for the ideal MHD normal modes of an expanding, solar wind-like plasma. In Section \ref{sec4}, we study an oblique mode of propagation for the waves, relative to a background magnetic field fixed in the direction of propagation. The dispersion relations are deduced and discussed in this section for this specific case. Finally, in Section \ref{sec5}, we summarize and conclude our results.

\section{MHD and EBM}
\label{sec2}

\begin{figure}
    \centering
    \includegraphics[width=1\linewidth]{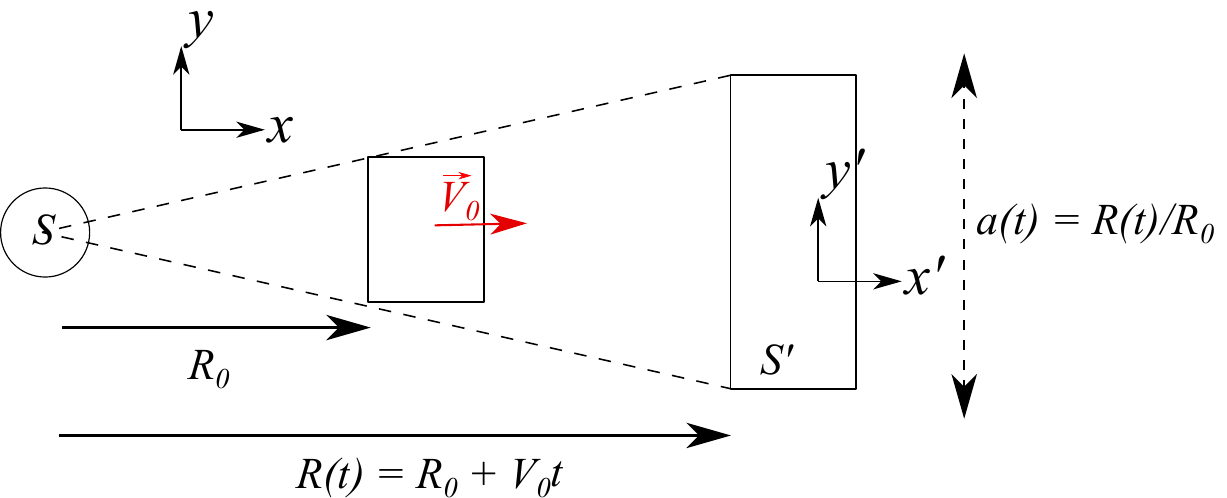}
    \caption{Cartesian approximation for a radially and spherically expanding plasma. The plasma box travels at a constant velocity $V_0$, and is observed from a reference system $S$ moving away at a distance $R(t)$. The EBM introduces a non-inertial reference frame $S'$ that moves along the plasma box at velocity $V_0$. Renormalization of the transverse coordinates by the expansion parameter $a(t)$ maintains the volume of the box constant.}
    \label{fig:1}
\end{figure}

In order to develop an analytical description of the linear normal modes of a collisionless magnetized plasma including expansion effects, we employ the EBM formalism described by Velli et al.~\cite{Velli1992}. The EBM allows us to mimic spherical plasma expansion through a Cartesian non-trivial change of coordinates to a non-inertial, co-moving frame. Given an inertial, fixed system of reference $S$ from which the plasma is expanding away, the EBM introduces a new system of reference $S'$ co-moving with the plasma at a constant velocity $V_0$. The position of the plasma box at time $t$ is given by $R(t) = R_0 + V_0 t$, where $R_0$ is the initial position. Figure \ref{fig:1} shows the Cartesian approximation and the relation between the reference systems $S$ and $S'$. Thus, the coordinates of both systems are related by a Galilean transformation in the radial direction $x$, and a renormalization in the transverse coordinates
\begin{align}
    x' = x-R(t), \qquad y'  = \frac{1}{a(t)}y,  \qquad z'=\frac{1}{a(t)} z,
\end{align}
\begin{align}
     a(t) = \frac{R(t)}{R_0} = 1 + \frac{U_0}{R_0}t.
\end{align}

Here, $a(t)$ is a dimensionless parameter that maintains the constant volume of the plasma box through the contraction of the transverse coordinates. This parameter enables the expansion of a plasma to be studied in a model that keeps its volume constant. This is because non-trivial temporal modifications that are explicitly dependent on this parameter will be introduced when the spatial and temporal derivatives are written in the co-moving frame. The spatial derivatives are
\begin{align}
    \frac{\partial}{\partial x} = \frac{\partial}{\partial x'}, \qquad \frac{\partial}{\partial y} = \frac{1}{a(t)} \frac{\partial}{\partial y'}, \qquad  \frac{\partial}{\partial z'} = \frac{1}{a(t)} \frac{\partial}{\partial z'}.
\end{align}
Following Grappin et al.~\cite{Grappin1993}, and using the same notation as Echeverría-Veas et al.~\cite{Echeverria-Veas2023}, the spatial gradients and temporal derivative can be written in the new system as
\begin{align}
    \nabla = \mathbb A^{-1}\cdot \nabla', \qquad \nabla_v = \mathbb A^{-1}\cdot \nabla_v', \qquad \frac{\partial}{\partial t} = \frac{\partial}{\partial t'} - \mathbf D \cdot \nabla',\label{eq4}
\end{align}
where
\begin{align}
    \mathbb A = \begin{pmatrix}
        1 & 0 & 0\\
        0 & a(t) & 0\\
        0 & 0 & a(t)
    \end{pmatrix},& \qquad \mathbf D = V_0 \left (\hat x + \frac{y'}{R(t)} \hat y +  \frac{z'}{R(t)} \hat z\right ),
    \\
    &\mathbf V_0 = V_0\left (\hat x + \frac{y}{R_0}\hat y +  \frac{z}{R_0}\hat z  \right ).
\end{align}

On the other hand, for the transformation of the flow velocity to the co-moving system, we can write $\mathbf u = \mathbb A \cdot \mathbf u' + \mathbf V_0$. Note that in the non-inertial reference frame $S'$, all the information about the expansion will be expressed explicitly through the $\mathbb A(t)$ tensor. Considering this, by using the previous transformations, it is possible to write the equations of ideal MHD in the expanding frame, neglecting the primes in the quantities, as (see Grappin et al.~\cite{Grappin1993} and Echeverría-Veas et al.~\cite{Echeverria-Veas2023} for a detailed derivation)
\begin{equation}
    \frac{\partial n}{\partial t} + \nabla \cdot (n \mathbf u) = - \frac{2 \dot a }{a} n,
    \label{eq7}
\end{equation}
\begin{equation}
    \frac{\partial p}{\partial t} + \mathbf u \cdot \nabla p + \gamma p \nabla \cdot \mathbf u = -  \gamma \frac{2 \dot a}{a} p,
    \label{eq8}
\end{equation}
\begin{equation}
     \frac{\partial \mathbf B}{\partial t} - [
            \nabla \times (\mathbf u \times \mathbf B)] = - \frac{\dot a}{a}\mathbb L \cdot \mathbf B,
            \label{eq9}
\end{equation}
\begin{equation}
    \frac{\partial \mathbf u}{\partial t} + (\mathbf u \cdot \nabla)\mathbf u + \frac{1}{8 \pi \rho} [(\mathbb A^{-1} \cdot \nabla )]B^2 - \frac{1}{4 \pi \rho} [\mathbf B \cdot  (\mathbb A^{-1}\cdot \nabla)] \mathbf B- \frac{1}{\rho}\nabla p = - \frac{\dot a}{a} \mathbb T \cdot \mathbf u\,,
    \label{eq10}
\end{equation}
where
\begin{align}
\mathbb L = \begin{pmatrix}
    2 & 0 & 0\\
    0 & 1 & 0\\
    0 & 0 & 1
\end{pmatrix}, \qquad \mathbb T = \begin{pmatrix}
    0 & 0 & 0\\
    0 & 1 & 0\\
    0 & 0 & 1
\end{pmatrix}.
\end{align}
Here, the momentum equation describes the bulk flow velocity of the plasma, which closely corresponds to the proton velocity due to the small electron-to-proton mass ratio. Note that, when writing the ideal MHD-EBM system of equations, we have neglected the Hall term in Equation \ref{eq9}, and assumed a scalar pressure equation with polytropic index $\gamma$ for Equation \ref{eq8}, thus closing the system of MHD equations within the EBM framework. An important detail to note is that, although the expansion of the solar wind is non-adiabatic as mentioned above, we use an adiabatic equation of state for the pressure. This is because the MHD waves we intend to study occur on sufficiently small timescales that we can safely assume there is no heat flux. This makes the adiabatic equation of state a good approximation. Following Echeverría-Veas et al.~\cite{Echeverria-Veas2023, Echeverria-Veas2024}, we have written the equations with the matrix $\mathbb A$ outside the definition of the gradients, as expressed in Equation \ref{eq4}. This allows us to explicitly identify the time dependencies introduced by the expansion in the equations. An important detail to take into account corresponds to the term $\dot a/a$ on the right-hand side of the equations. In the literature, this has usually been identified as a time scale of the expansion, where $\dot a/a = 1/ \tau$, writing $\tau$ as the expansion time scale~\cite{Nariyuki2015,Shi2020}. Equations \eqref{eq7}--\eqref{eq10} are commonly found in the literature when performing computer simulations of expanding plasma phenomena within the EBM framework~\cite{Grappin1996,Liewer2001,Ofman2011}. 

\section{MHD-EBM Linear Plasma Waves}
\label{sec3}

Following the standard procedure in the literature~\cite{Alfven1942}, MHD waves are studied in a linear regime by considering small-amplitude perturbations of macroscopic plasma quantities, described by the system of ideal MHD equations for a perfectly conducting, infinitely extended plasma. In this linear framework, the quantities $n$, $p$, $\mathbf B$, and $\mathbf u$ are decomposed into a background and a small perturbation. It is worth noting, however, that some MHD wave modes—such as the Alfvén wave—also admit exact nonlinear solutions under certain conditions.
\begin{align}
    n = n^{(0)} +  n^{(1)},\qquad p = p^{(0)} + p^{(1)}, \qquad \mathbf B = \mathbf B^{(0)} +  \mathbf B^{(1)}, \qquad \mathbf u = \mathbf u^{(0)} + \mathbf u^{(1)},
\end{align}
where in the linear approximation, for each quantity $\phi$, we require that $\phi^{(0)} \gg \phi^{(1)}$. Our solution for the traveling-wave perturbation is given by $\phi^{(1)}(t) = \phi_1 e^{i (\mathbf k\cdot \mathbf r - \omega t)}$, meaning that waves can propagate in any direction. Unlike the standard non-expanding procedure in the literature, the background quantities will not be spatially homogeneous and time-independent, due to the action of inertial forces on the EBM frame.  We aim to write the solutions for $n_1$, $p_1$, and $\mathbf B_1$, and insert them into the first-order momentum equation. This will yield a wave equation whose roots correspond to the expanding dispersion relations $\omega (k, R)$.

\subsection{Zero-order Equations}

The resulting linearized MHD-EBM equations for the background quantities are 
\begin{equation}
    \frac{\partial n^{(0)}}{\partial t} + \nabla \cdot (n^{(0)} \mathbf{u}^{(0)}) = - \frac{2 \dot a}{a} n^{(0)},
\end{equation}
\begin{equation}
    \frac{\partial p^{(0)}}{\partial t} + \mathbf u^{(0)} \cdot \nabla p^{(0)} + \gamma p^{(0)} \nabla \cdot \mathbf u^{(0)} = - \gamma \frac{2 \dot a}{a} p^{(0)},
\end{equation}
\begin{equation}
        \frac{\partial \mathbf B^{(0)}}{\partial t} - [\nabla \times (\mathbf u^{(0)} \times \mathbf B^{(0)})] = -\frac{\dot a}{a} \mathbb L \cdot \mathbf{B}^{(0)},
\end{equation}
\begin{equation}
\begin{aligned}
        (n^{(0)} m_p) \frac{\partial \mathbf u^{(0)}}{\partial t} + (n^{(0)} m_p) (\mathbf{u}^{(0)} \cdot \nabla) \mathbf{u}^{(0)} + \nabla p^{(0)} +  &\\\frac{1}{8 \pi} [(\mathbb A^{-1} \cdot \nabla) B^{(0)2}] - \frac{1}{4\pi} [B_0 \cdot (\mathbb A^{-1} \cdot \nabla) B_0]& = - (n^{(0)} m_p) \frac{\dot a}{a} \mathbb T \cdot \mathbf u^{(0)}.
\end{aligned}
\end{equation}
The mass density of the fluid has been linearized such that $\rho = n m_p$, then $\rho = m_p (n^{(0)} +  n^{(1)})$. Assuming a background fluid velocity $\mathbf u^{(0)} = 0$, the resulting differential equations can be easily solved for each quantity, obtaining the background profiles as 
\begin{equation}
    n^{(0)} = \frac{n_0}{a^2},\\
\end{equation}
\begin{equation}
    p^{(0)} = \frac{n_0}{a^{2 \gamma}},\\
\end{equation}
\begin{equation}
    \mathbf B^{(0)} = \mathbb Z^{-1} \cdot \mathbf B_0,\\
\end{equation}
where
\begin{equation}
    \mathbb Z = \begin{pmatrix}
        a(t)^2 & 0 & 0\\
        0 & a(t) & 0\\
        0 & 0 & a(t)
    \end{pmatrix}.
\end{equation}
The obtained profiles of the background quantities are consistent with those of Grappin et al.~\cite{Grappin1993}. Furthermore, since $a(t) \propto R(t)$, the shape of the radial decay of the magnetic field shows rotation in the x-y plane, naturally representing the Parker spiral.

\subsection{First-order Equations}

We can write the resulting system of MHD-EBM equations for the first-order perturbation. Maintaining $\mathbf u^{(0)} = 0$, the equations follow as
\begin{equation}
    \frac{\partial n^{(1)}}{\partial t}  + \nabla \cdot (n^{(0)} \mathbf u^{(1)}) = - \frac{2 \dot a}{a} n^{(1)},
    \label{eq21}
\end{equation}
\begin{equation}
        \frac{\partial p^{(1)}}{\partial t} +  \mathbf u^{(1)} \cdot \nabla p^{(0)} + \gamma p^{(0)} \nabla \cdot \mathbf u^{(1)}  = - \frac{2 \gamma V_0}{a R_0} p^{(1)},
\end{equation}
\begin{equation}
        \frac{\partial \mathbf B^{(1)}}{\partial t} + [\nabla \times (\mathbf u^{(1)} \times \mathbf B^{(0)})]  = -\frac{\dot a}{a} \mathbb L \cdot \mathbf{B}^{(1)},
\end{equation}
\begin{equation}
\begin{aligned}
        &(n^{(0)} m_p) \frac{\partial \mathbf u^{(1)}}{\partial t} + \nabla p^{(1)} + \frac{1}{4 \pi} [(\mathbb A^{-1} \cdot \nabla) \mathbf B^{(0)} \mathbf B^{(1)}] + \frac{1}{4 \pi} [(\mathbb A^{-1} \cdot \nabla) \mathbf B^{(1)} \mathbf B^{(0)}]  -\\&  \frac{1}{4\pi} [\mathbf B^{(0)} \cdot (\mathbb A^{-1} \cdot \nabla) \mathbf B^{(1)}] - \frac{1}{4\pi} [\mathbf B^{(1)} \cdot (\mathbb A^{-1} \cdot \nabla) \mathbf B^{(0)}]  = - (n^{(0)} m_p) \frac{\dot a}{a} \mathbb T \cdot \mathbf u^{(1)}.\label{eq24}
\end{aligned}
\end{equation}

\begin{figure}
    \centering
    \includegraphics[width=0.4\linewidth]{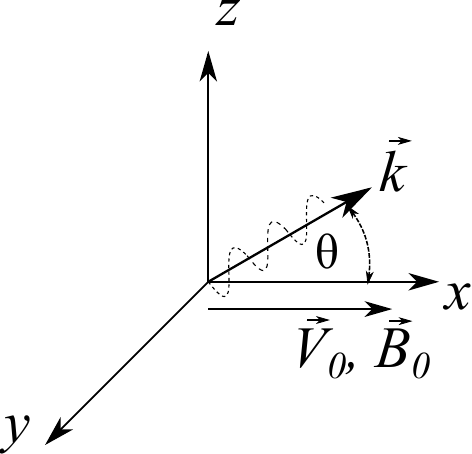}
    \caption{Propagation diagram for MHD waves in this scenario. Here, $x$ is the radial direction from which the plasma expands at velocity $V_0$. Thus, the background magnetic field is aligned with the direction of propagation, and the wave vector $k$ traces an angle $theta$ in the x-z plane to the direction of propagation.}
    \label{fig:2}
\end{figure}

By inserting the zeroth-order solutions for density, pressure, and magnetic field into the first-order equations, we can solve for the perturbed quantities. For Equation \eqref{eq21}, we can write
\begin{equation}
    \frac{\partial n^{(1)}}{\partial t} + \frac{2 \dot a}{a} n^{(1)} = - \nabla \cdot \left (\frac{n_0}{a^2} \mathbf u_1 \right ).
\end{equation}
It is clear that, by multiplying both sides by $a(t)^2$, a product rule appears at the left-hand side of the equation. This allows us to write the equation to be solved as
\begin{equation}
    \frac{d}{dt} (n^{(1)} a^2) = - \nabla \cdot \left ({n_0} \mathbf u_1 \right ).
\end{equation}
This equation can be solved using the standard Fourier analysis. In this analysis, the operators $\nabla$ and $\frac{d}{dt}$ can be written as$\nabla \to i \mathbf k$ and $\frac{d}{dt} \to - i \omega$, respectively, based on the solution given for the traveling-wave solutions of the perturbations. However, transforming the term $a(t)^2$ to Fourier space using the product rule complicates this analysis, as chain rules would appear, making this a differential equation for $\omega$. In a weak expansion limit, where the frequency scale $1/\tau$ of the solar wind expansion is much larger than that of the oscillations $\omega$, it is possible to approximate the total derivative in such a way that it does not affect the $a(t)^2$ term, given the larger time scale. Thus, the expansion does not affect the traveling-wave perturbations. Hence, as $n^{(1)} = n_1 e^{i(\mathbf k \cdot \mathbf r - \omega t)}$ and $u^{(1)} = u_1 e^{i(\mathbf k \cdot \mathbf r - \omega t)}$ , we obtain
\begin{equation}
    n_1 = \frac{(\mathbf k \cdot \mathbf u_1) n_0}{a^2 \omega}.
    \label{eq27}
\end{equation}
Thus, we obtain a result that can be inserted into the linearized momentum equation to derive an MHD-EBM wave equation. This result is similar to those known in the literature, but with the addition of a $1/a^2$ dependence from the radial decrease of the background quantities. In this case, we are solving for a situation in which the background quantities, which extend throughout space, are affected by plasma expansion. However, the perturbed quantities, which oscillate at a much higher frequency than the expansion frequency, are not affected. For the pressure and magnetic field equation, multiplying both sides by $a^{2 \gamma}$ and $\mathbb Z$ respectively, we obtain
\begin{equation}
    p_1 = \frac{\gamma p_0 \mathbf k \cdot \mathbf u_1}{\omega a^2\gamma},
    \label{eq28}
\end{equation}
\begin{equation}
    \mathbf B_1 = - \frac{1}{\omega} [\mathbf k \times (\mathbf u_1 \times (\mathbb Z^{-1} \cdot \mathbf B_0))].
    \label{eq29}
\end{equation}
Inserting Equation \eqref{eq27}, \eqref{eq28}, and \eqref{eq29} into Equation \eqref{eq24} using the same procedure as before, the wave equation can be written from Equation \eqref{eq24} as
\begin{equation}
    \omega^2 \mathbf {u_1} =   \frac{1}{4\pi n_0 m_p} (\mathbf k  \times [\mathbf k \times (\mathbf{u_1} \times \mathbb \ \mathbf B_0 )]  ) \times (\mathbb Z^{-1} \cdot \mathbf B_0 ) + \frac{\gamma p_0}{m n_0} a^{2 - 2 \gamma}\mathbf k (\mathbf k \cdot \mathbf{u_1}),\label{eq30}
\end{equation}
where vector identities have been used to write the cross products. From here, we can consider the different propagation scenarios for the waves with respect to the background field $\mathbf B_0$, and thus solve the cross wave equation. For oblique propagation, we consider the simple case where the plasma box propagates in the same direction as the background magnetic field at $\hat x$, and the wave vector forms an angle $\theta$ with the magnetic field and the direction of propagation. Figure \ref{fig:2} illustrates this model. This scenario is a compromise, given that the direction of the solar magnetic field varies across the heliosphere~\cite{Adrian2016}. Nevertheless, this is a reasonable approximation for comparing the results with those existing in the literature for the non-expanding case. Once the cross products have been solved under this scenario, it is possible to determine the expanding dispersion relations $\omega(k,R)$ from the momentum equation.

  \section{Results: Oblique Propagation}
\label{sec4}

We now study the propagation of oblique waves to the background field $B_0$. The momentum $\mathbf u_1$, background magnetic field $\mathbf B_0$ and wave vector $\mathbf k$ can thus be defined in this scenario as
\begin{equation}
    \mathbf {{u_1}}  = (u_x, u_y,  u_z),
    \label{eq31}
\end{equation}

\begin{equation}
    \mathbf B_0 = (B_0, 0, 0), \qquad  \mathbb Z^{-1} \cdot \mathbf B_0  = (B_0/a^2,0, 0),   
    \label{eq32}
\end{equation}

\begin{equation}
    \mathbf k  = (k\cos \theta,0,k \sin \theta).
    \label{eq33}
\end{equation}

Inserting Equations \eqref{eq31}-\eqref{eq33} into Equation \eqref{eq30}, then dividing by $k^2$ and factoring out $B_0^2$, we obtain the wave equation in matrix form as

\begin{equation}
    \begin{pmatrix}
    \omega^2/k^2 - \tilde v_S(R)^2 \cos^2 \theta & 0 & - \tilde v_S(R)^2 \sin \theta \cos \theta\\
    0 & \omega^2/k^2 - \tilde v_A (R)^2 \cos^2 \theta & 0\\
    - \tilde v_S(R)^2 \sin \theta \cos \theta & 0 & \omega^2/k^2 - \tilde v_s (R)^2 \sin^2 \theta - \tilde v_A (R)^2 
\end{pmatrix} \begin{pmatrix}
    u_x \\ u_y \\ u_z
\end{pmatrix} = 0.\label{eq34}
\end{equation}
In the previous expressions, the Alfvén and sound speeds have been rewritten to implicitly express their radial dependence. Thus, the modification introduced by the expansion of the MHD waves corresponds to the radial decrease of these two speeds, where
\begin{equation}
    \tilde v_A(R)^2 = \frac{B_0^2}{4 \pi n_o m a^2} = v_A^2\frac{R_0^2}{R^2},
\end{equation}
\begin{equation}
    \tilde v_S(R)^2 = \frac{\gamma p_0}{n_0 m} a^{2 - 2 \gamma} =v_S^2  \frac{R_0^{2\gamma - 2}}{R^{2\gamma  - 2}},
\end{equation}
where we have used $a(t) = R/R_0$ to express the decay of the speeds as a function of heliocentric distance $R$. Here, $v_A$ and $v_S$ correspond to the Alfvén and sound speeds, respectively, in the non-expanding plasma scenario. Equation \eqref{eq34} has non-trivial solutions when the determinant of the left-hand matrix is zero, which yields the dispersion relation for expanding MHD waves as
\begin{equation}(\omega^2/k^2 - \tilde  v_A(R)^2 \cos \theta) ((\omega/k)^4 - \omega^2 (\tilde v_A(R)^2 + \tilde v_S(R)^2 )/k^2 + \tilde v_A(R)^2 \tilde v_S(R)^2 \cos^2 \theta) = 0.\label{eq37}\end{equation}
This dispersion relation has the same form as the ideal, non-expanding MHD, but with explicit radial dependencies for Alfvén and sound speeds. In fact, for the non-expanding case given by $a(t) = 1$, we recover the well-known ideal MHD dispersion relation; namely
\begin{equation}(\omega^2/k^2 - v_A^2 \cos \theta) ((\omega/k)^4 - \omega^2 ( v_A^2 +  v_S^2 )/k^2 +  v_A^2 v_S^2 \cos^2 \theta) = 0.\end{equation}
 Equation \eqref{eq37} has three eigenfrequencies, corresponding to the three normal modes of the expanding ideal MHD system:
\begin{equation}
    \omega^2 = \frac{k^2}{2} \left ( \tilde v_A(R)^2 + \tilde v_S(R)^2 \right ) + \frac{k^2}{2} \left[ \left ( \tilde v_A(R)^2 - v_S(R)^2 \right )^2 + 4  \tilde v_A(R)^2 \tilde v_S(R)^2 \sin^2 \theta \right ]^{1/2},\label{eq38}
\end{equation}
\begin{equation}
    \omega^2 =  \frac{k^2}{2} \left ( \tilde v_A(R)^2 + \tilde v_S(R)^2 \right ) - \frac{k^2}{2} \left[ \left ( \tilde v_A(R)^2 - v_S(R)^2 \right )^2 + 4  \tilde v_A(R)^2 \tilde v_S(R)^2 \sin^2 \theta \right ]^{1/2},\label{eq39}  
\end{equation}
\begin{equation}
    \omega^2 = k^2 \tilde v_A(R)^2  \cos^2 \theta.\label{eq40} 
\end{equation}
Equations \eqref{eq38}, \eqref{eq39}, and \eqref{eq40} represent the fast magnetosonic mode, the slow magnetosonic mode, and the transverse Alfvén mode, respectively.  Comparing them to the well-known non-expanding case shows that incorporating plasma expansion into the ideal MHD equations via the EBM framework results in dispersion relations with radial decreases for the Alfvén and sound speeds. It is worth noting that we recover the usual linear MHD waves dispersion relations in the non-expanding limit ($a(t) = R/R_0 = 1$). The dispersion relation for the transverse Alfvén wave, as described by Equation \eqref{eq40}, predicts a decrease in wave frequency with respect to heliocentric distance in the form of $\omega \propto  1/R$. This coincides with Parker's linear prediction of the energy decay of Alfvén waves in an expanding plasma.~\cite{Parker1965}. In our model, this frequency decrease is associated with the radial decrease of the Alfvén velocity, as $\tilde v_A(R) \propto 1/R$. Conversely, for both the fast and slow magnetosonic modes, frequency decay is proportional to the polytropic index $\gamma$, as the plasma sound velocity $\tilde v_S(R)$ explicitly depends on this parameter, as $\tilde v_S(R) \propto 1/R^{\gamma - 1}$ in the expanding frame. This index does not have a unique value for all solar wind extents and scales because it characterizes specific thermodynamic processes at different scales and plasma regimes. Therefore, it may have different values in the same environment and vary throughout the heliosphere~\cite{Livadiotis2023}.

Since our results for the wave frequency $\omega(k,R)$ depend explicitly on the value of the polytropic index, we considered three models regarding the evolution of this index throughout solar wind expansion. The first model corresponds to a quasi-adiabatic approximation in which $\gamma$ has a constant value of $\gamma = 5/3$ throughout the expansion. For the second scenario, we use the model proposed by Livadiotis et al.~\cite{Livadiotis2021} for the radial profile of the polytropic index $\gamma$ of the solar wind throughout the heliosphere. In this model, the index is approximated as an adiabatic one ($\gamma = 5/3$) in the inner heliosphere, and begins to decrease from subadiabatic indices until near-zero values from $\sim 20$ AU. Finally, we consider the polytropic index model proposed by Nicolaou et al.~\cite{Nicolaou2020}, in which the value of the polytropic index depends on the ratio $\delta q/\delta w$ of the energy supplied to the system as heat to the energy supplied as work, and is given by
\begin{equation}
    \gamma = \frac{2}{f} \left ( 1 - \frac{\delta q}{\delta w}\right ) + 1,
    \label{eq42}
\end{equation}
where $f$ corresponds to the degrees of freedom of the system, having $f = 3$ in this case. Since the Alfvén wave dispersion relation does not depend on the polytropic index and the slow magnetosonic wave is not noticeably affected by the choice of the model, we present the results for the two magnetosonic modes in Subsection \ref{sub4.1}, considering a polytropic index value that evolves radially according to the Livadiotis et al.~\cite{Livadiotis2021} model. However, since the fast magnetosonic wave is significantly affected by the evolution and choice of the polytropic index, in Subsection \ref{sub4.2} we analyze the dispersion relation for the fast magnetosonic mode for different models of the polytropic index. 
\begin{figure}
    \centering
    \includegraphics[scale = 0.75]{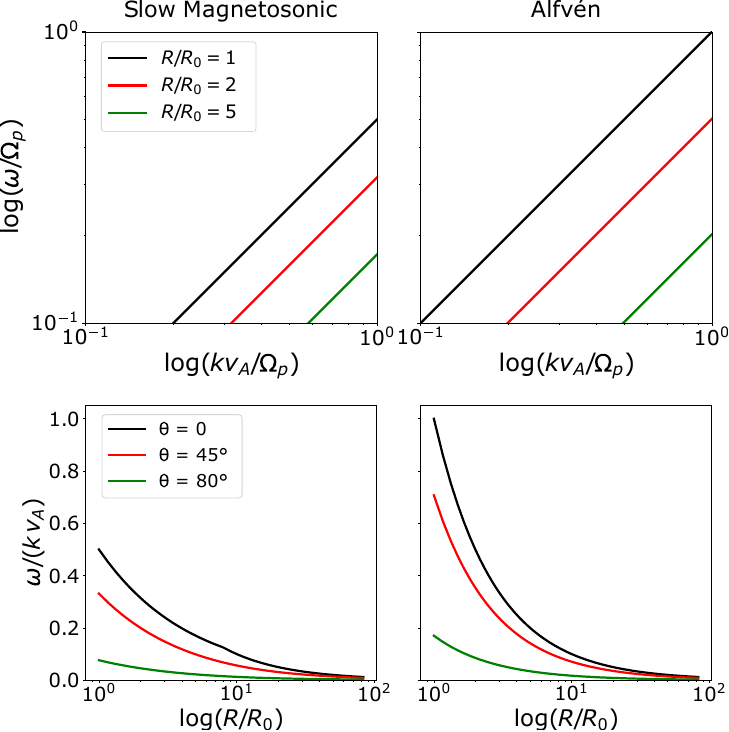}
    \caption{(Top) Normalized dispersion relations for each mode with $\theta$ fixed at 0$^\circ$, for different values of the $R/R_0$ ratio. Here, $R_0 = 0.3$~AU, and $\Omega_p$ is the proton gyrofrequency. (Bottom) Normalized phase speeds of the waves for various $\theta$ values, as functions of heliocentric distance.}
    \label{fig:3}
\end{figure}

\subsection{Alfvén and Slow Magnetosonic Wave}
\label{sub4.1}

Figure \ref{fig:3} (top) shows the obtained normalized expanding dispersion relations that incorporate the radial decrease for the polytropic index of the solar wind, $\gamma$. Here, we present $\omega/\Omega_p$ as a function of $k v_A/\Omega_p$ for different values of the expansion parameter $a(t) = R/R_0$.  Both axes have been adjusted to a logarithmic scale. Here, $\Omega_p$ is the proton gyrofrequency, $R_0 = 0.3$ AU, and for the non-expanding plasma speeds, $v_A/v_S = 2$. $R/R_0 = 1$ (black line) represents the non-expanding case, recovering the ideal MHD solution where the normalized solution for the Alfvén wave corresponds to the identity function. As the solar wind expands, the frequency of the slow magnetosonic and Alfvén modes decays with radial heliocentric distance, as deduced analytically before. 

To better illustrate this phenomenon, Figure \ref{fig:3} (bottom) shows the radial evolution of the normalized phase speed ($\omega/(k v_A)$) of each mode at propagation angles $\theta = 0^\circ, 45^\circ$ and near $90^\circ$. The x-axis, which represents normalized radial distance $(R/R_0)$, has been adjusted logarithmically for better visualization. Both the slow magnetosonic mode and the Alfvén mode decrease their phase speed as the plasma expands. As expected, the phase speed of the Alfvén wave is initially larger than that of the slow magnetosonic wave; however, it decreases rapidly with heliocentric distance. This can be traced back directly to our theoretical result for the dispersion relation of the mode. Our results suggest that, at sufficiently large heliocentric distances, the frequency of these waves may be too low for detection.

\subsection{Fast Magnetosonic Wave}
\label{sub4.2}

\begin{figure}[!h]
    \centering
    \includegraphics[scale = 0.35]{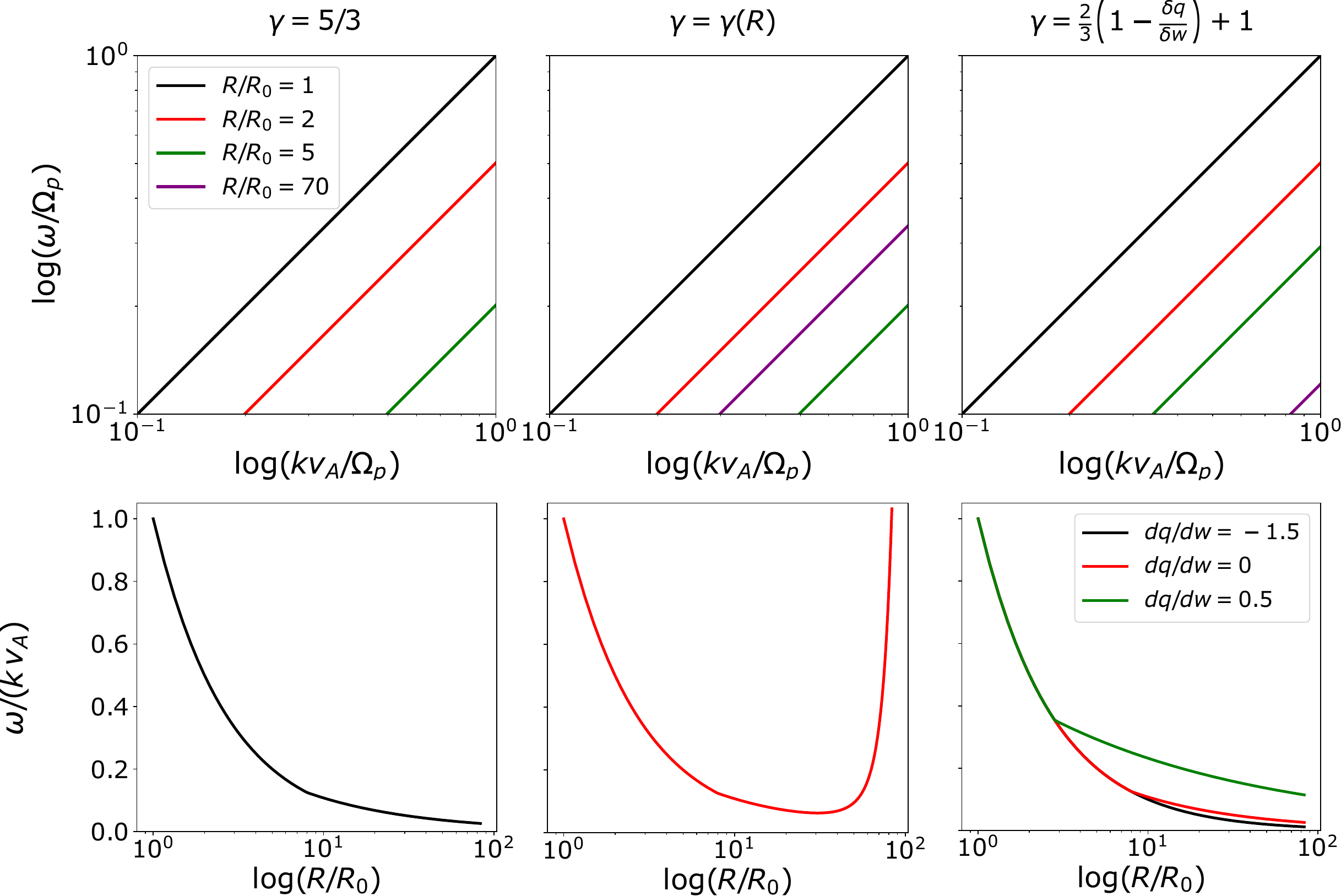}
    \caption{(Top) Normalized dispersion relations for each mode with $\theta$ fixed at 0$^\circ$. Here, $R_0 = 0.3$AU, and $\Omega_p$ is the proton gyrofrequency. (Bottom) Normalized phase speeds of the waves for various $\theta$ values, as functions of heliocentric distance.}
    \label{fig:4}
\end{figure}

In order to study the radial evolution of the fast magnetosonic mode throughout the solar wind expansion, we have incorporated three models to determine the evolution of the polytropic index $\gamma$ as the plasma expands. As mentioned, the first model consists of a quasi-adiabatic approximation such that $\gamma$ is constant at $\gamma = 5/3$. In the second model, the value remains constant at $\gamma = 5/3$ up to approximately 20 AU. Then, it begins to decrease to 0 in the outer heliosphere~\cite{Livadiotis2021}. The third model depends on the ratio of the energy supplied to the plasma by heat to the energy supplied by work $(\delta q/\delta w)$, as given by Equation \eqref{eq42}. Figure \ref{fig:4} (top) shows the normalized wave frequency $\omega/\Omega_p$ as a function of the normalized wave number $k v_A/\Omega_p$ at different heliocentric distances $R/R_0$. Here, $\theta = 0$ in all cases. For the third model, $\delta q/\delta w = 0.5$, which represents the presence of mechanisms supplying heat to the expanding plasma.

If the polytropic index $\gamma$ is constant at $\gamma = 5/3$, a decrease of the wave frequency with heliocentric distance is evident, becoming significantly smaller than close to the Sun in the outer heliosphere ($R/R_0 = 70$, or $R = 20$ AU).  The wave frequency then monotonically decreases with distance, in a similar manner to the slow magnetosonic and Alfvén modes.
On the other hand, when considering the second model in which the gamma polytropic index decreases to zero in the outer heliosphere, the wave frequency initially decreases with heliocentric distance, but then increases again in the outer heliosphere where $R/R_0 = 70$. Since the left side of the dispersion relation is proportional to $\omega \propto R^{\gamma - 1}$, the wave will start to increase its frequency and thus phase speed when $\gamma < 1$. Observations have shown that this is indeed the case for the polytropic index in the outer heliosphere~\cite{Elliott2019}. 
Finally, we examine the third model with $dq/dw = 0.5$. A positive value of this ratio indicates the presence of plasma heating mechanisms, something expected during the non-adiabatic expansion of the solar wind. In this case, the wave frequency decreases with heliocentric distance, though more slowly than in the first case, without showing acceleration. 

Figure \ref{fig:4} (bottom) shows the evolution of the normalized phase speed ($\omega/(k v_A)$) according to the heliocentric distance $R/R_0$, for each mode at a fixed propagation angle of $\theta = 0^\circ$. Same as before, $R_0 = 0.3$ AU, and $v_A/v_S = 2$. Following the line of what was previously examined for the wave frequency,  the use of the first, adiabatic model evidences a radial decrease in phase speed, approaching zero in the outer heliosphere. The second model of radial decrease for the index evidences an acceleration of the phase speed, from the theoretical prediction of frequency evolution with $1/R^{\gamma - 1}$. Therefore, the phase speed starts to increase when $\gamma < 1$, starting at approximately 20 AU. To examine the third model, we considered three different values of the $\delta q/\delta w$ ratio.  The case of $\delta q/\delta w = 0.5$ implies the presence of mechanisms heating the plasma; therefore, the frequency decreases more slowly than in the adiabatic case of $\delta q/\delta w = 0$.  Finally, the case of $\delta q/\delta w = -.15$ implies the presence of mechanisms retaining heat from the plasma. This causes the wave frequency to decrease slightly faster, though not considerably.

\section{Discussion and Conclusions}
\label{sec5}

In this work, we have derived the dispersion relations of the three normal modes of ideal MHD in an expanding frame given by the EBM. Our results show that the modification of the dispersion relations due to plasma expansion manifests itself in the radial decay of the plasma speeds (Alfvén and sound), which can be traced back to the radial profiles obtained for the plasma background quantities. The magnetic field, density, and pressure are modified during the radial expansion of the plasma due to the inertial forces inherent to the transverse coordinate contraction in the EBM frame. This relates to the hydrodynamic strain-rate tensor, which characterizes a fluid's stretching and vorticity rate~\cite{Batchelor2000}. Thus, coordinate contraction and stretching in the EBM frame translate to fluid stretching in the MHD-EBM frame. This modifies the plasma background quantities. This is easily observable in the case of the background magnetic field, where decay is faster in the uncompressed propagation direction than in the re-normalized transverse directions. This directly translates when writing the Alfvén and sound speeds in the expanding frame, thus modifying the dispersion relations of linear MHD waves. This result can be directly interpreted for the Alfvén and slow magnetosonic modes, which decrease radially until they almost vanish in the outer heliosphere. 

The slow magnetosonic mode is not appreciably affected by the choice for the evolution of the polytropic index $\gamma$, because the other terms in the dispersion relation predominate over the dependence on $1/R^{\gamma - 1}$. However, the radial profile of the dispersion relation for the fast magnetosonic mode is significantly affected by the choice of $\gamma$. When the polytropic index approaches a quasi-adiabatic value $(\gamma = 5/3)$, the fast magnetosonic mode exhibits a similar radial decay profile to the other two modes, almost vanishing in the outer heliosphere. However, if the polytropic index is chosen to be sub-adiabatic as in the third model $(\delta q/\delta w = 0.5)$, this radial decay is slower. In this case, non-zero values are found for the frequency and phase speed in the outer heliosphere. Nevertheless, in the second model where $\gamma$ decays with radial distance in the outer heliosphere, the decay of the index to sub-adiabatic values causes the fast magnetosonic mode to increase in frequency after around 20 AU, and therefore in phase speed, when $\gamma < 1$. This acceleration may be due to the thermodynamic evolution of the plasma becoming a subadiabatic process in the outer heliosphere. Alternatively, it may be solely a result of the idealization of the conditions studied in the ideal magnetohydrodynamic (MHD) framework. More theoretical and observational research is needed to better understand this phenomenon. However, an alternative approach can be developed based on this result. Since deducing the dispersion relations in the expanding frame provides information about the ideal MHD waves and their properties at different heliocentric distances, this method can be used backwards to calculate the polytropic index $\gamma$, at different heliocentric distances, having observational data about the waves. 

It is worth noting that the radial decay of phase speeds, and thus wave frequency, along with plasma expansion, quantifies a loss of wave energy during expansion. However, when we translate these scenarios to a system such as the solar wind, the picture of wave propagation along the plasma expansion becomes much more complex. Nonlinear interactions, large amplitude waves, turbulence at different scales, and coupling between different wave modes dominate~\cite{Bruno2013,Marino2023}. Nevertheless, we pose this problem as the beginning of a more general study of wave propagation and scattering in an expanding medium. Generalizations closer to the sun could include using an expanding non-adiabatic pressure equation and an arbitrary background magnetic field direction relative to wave propagation. However, this is beyond the scope of the present study. In any case, the results of this study can be applied to future theoretical investigations of environments such as the expanding solar wind. While this study focuses on a simplified configuration in which the background magnetic field is aligned with the direction of expansion, the presented methodology can be extended to more general scenarios, including oblique magnetic field geometries or anisotropic background conditions. Such generalizations would provide a more general framework for comparing theoretical predictions with in situ measurements of the solar wind at various heliocentric distances.

\ack
We are grateful for the support of ANID Chile through the National Doctoral Scholarships No. 21250323 (SS), and FONDECYT grants No. 1240281 (PSM) and 1230094 (FAA). We would also like to thank the Organizing Committee of the 17th Latin American Workshop on Plasma Physics LAWPP-2025, held in Santiago, Chile, for a great and fruitful scientific meeting, in which an early version of this work was presented.

\section*{References}

\bibliographystyle{iopart-num}
\bibliography{bibliography}

\end{document}